\documentclass[bibyear]{aa}  
\usepackage{graphicx}
\usepackage{txfonts}
\usepackage{enumerate}
\usepackage{txfonts}
\usepackage{graphicx}

\begin{document}

\title{A collection of model stellar spectra for spectral types B to early-M} 
  
\author{C. Allende Prieto\inst{1,2}
  \and L. Koesterke\inst{3} 
  \and I. Hubeny\inst{4}
  \and M. A. Bautista\inst{5}
  \and P. S. Barklem\inst{6}
  \and S. N. Nahar\inst{7}
  }

   \institute{Instituto de Astrof\'{\i}sica de Canarias,
              V\'{\i}a L\'actea, 38205 La Laguna, Tenerife, Spain\\              
         \and
             Universidad de La Laguna, Departamento de Astrof\'{\i}sica, 
             38206 La Laguna, Tenerife, Spain \\
         \and
             Texas Advanced Computing Center, The University of Texas at Austin, Austin, TX 78759, USA \\
         \and
             Steward Observatory, University of Arizona, 933 N. Cherry Ave., Tucson, AZ, 85721, USA \\
         \and
         Department of Physics, Western Michigan University, Kalamazoo, MI 49008, USA \\
         \and
         Theoretical Astrophysics, Department of Physics and Astronomy, Uppsala University,
         Box 516, SE-751 20 Uppsala, Sweden \\
         \and
         Department of Astronomy, The Ohio State University, Columbus, OH 43210, USA
             }

%                        \email{callende@iac.es}
%
       %      \thanks{The university of heaven temporarily does not
       %              accept e-mails}

   \date{submitted December 18, 2017; accepted June 29, 2018}

% \abstract{}{}{}{}{} 
% 5 {} token are mandatory
 
  \abstract{
  Models of stellar spectra are necessary for interpreting  
  light from individual stars, planets, integrated stellar populations, nebulae,  
  and the interstellar medium.}
  {We provide a comprehensive and homogeneous collection of synthetic spectra 
  for a wide range of atmospheric parameters and chemical compositions.}
  {We compile atomic and molecular data from the literature.
  We adopt the largest and most recent set of ATLAS9 model atmospheres, and 
  use the radiative code ASS$\epsilon$T.}
  {The resulting collection of spectra is made publicly available at medium and 
  high-resolution ($R\equiv\lambda/\delta\lambda = 10,000$, 100,000 and  300,000) spectral grids,
  which include variations in effective temperature between 3500 K and 30,000 K, 
  surface gravity ($0\le \log g \le 5$), and metallicity ($-5 \le$[Fe/H]$\le +0.5$), spanning
  the wavelength interval 120-6500 nm. 
  A second set of denser grids with additional dimensions, [$\alpha$/Fe] and micro-turbulence, 
  are also provided (covering 200-2500 nm). We compare models with observations for a 
  few representative cases.}{}
   

\authorrunning{Allende Prieto et al.}
\titlerunning{Model Stellar Spectra} 

   \maketitle
%
%________________________________________________________________

\section{Introduction}
\label{intro}

The success of the theory of model atmospheres and radiation transfer 
brought us, in the 1970s, collections of synthetic spectra that resembled
quite closely the spectra of real stars (see, e.g., Gustafsson et al. 1975, Kurucz 1979).
Progress since has been significant regarding the quantity and quality of
the necessary atomic and molecular data (see, e.g., Hummer et al. 1993;  Seaton et al. 1994; 
Goorvitch 1994; Barklem \& O'Mara 1998; 
Barklem et al. 1998, 2000; Badnell et al. 2005). 

In the last decades, remarkable progress has been made relaxing the most limiting 
approximations adopted in the construction of classical model atmospheres. 
Departures from local
thermodynamical equilibrium (LTE) are now being routinely considered in models for
hot stars (e.g., Hubeny \& Lanz 1995; Puls et al. 1996; Hillier 2012), and their 
effects on level populations in cooler stars has been carefully assessed for 
a wide range of ions (see, e.g., Korn 2008, Asplund et al. 2009). Hydrodynamics 
are also included in some cases, 
but the fact that such models are necessarily three-dimensional and time-dependent 
has limited their production and distribution (Trampedach et al. 2013; Tremblay et al. 2013). 
The first library with synthetic spectra for nearly 100 three-dimensional model atmospheres is
now available (Ludwig et al. 2018), but many practical applications require much
finer grids, which have, at least at the present time, to be computed with hydrostatic models.

To meet our own needs, and those of the community in a large number 
of ongoing and upcoming spectroscopic surveys (see, e.g., Hutchinson et al. 2016, 
Recio-Blanco et al. 2016, or Starkenburg et al. 2017), we have decided
to take a snapshot of the data and codes currently available to us to produce 
a collection of model stellar spectra. The majority of the ingredients required
to compute stellar spectra, including
atomic and molecular data, physical approximations, algorithms, and 
computer codes, are constantly being improved, and many of the 
adopted  inputs and tools are already outdated. We nevertheless consider
that the calculations presented here represent an improvement over other
collections available, and will be useful for multiple applications. Examples 
of other model libraries publicly available are those by Palacios et al. (2010;
based on the Gustafsson et al. (2008) MARCS models, see also van Eck et al. 2017), 
Husser et al. (2013; based on the Phoenix code of Hauschildt et al. 1997), 
Coelho (2014), or Bohlin et al. (2017). Our library includes higher temperatures 
than those available in the MARCS or PHOENIX collections, and more recent and accurate 
continuum opacities than those based on Kurucz's SYNTHE code.

Section \ref{ma} provides a brief description of the adopted model atmospheres.
Sections \ref{cont} and \ref{line} describe the calculation of opacities and
Section \ref{rt}  describes the spectral synthesis calculations. The span and limitations of
the grids of stellar spectra are discussed in Section \ref{grids}. Section \ref{data}
illustrates a limited comparison with observations, 
while we provide a brief summary in Section \ref{sum}.

\section{Model ingredients and computations}

\subsection{Model atmospheres}
\label{ma}

We have adopted the most recent collection of ATLAS9 model atmospheres published
by M\'esz\'aros et al. (2012). These models were computed with the public and well-known
code by Kurucz (1979, and subsequent updates), which uses opacity distribution
functions to handle line absorption. This collection\footnote{Available on the
web from www.iac.es/proyecto/ATLAS-APOGEE/.} includes more than a
million model atmospheres originally computed for the Apache Point Observatory Galactic Evolution
Experiment (APOGEE; Majewski et al. 2017).
A similar collection of MARCS models (Gustafsson et al. 2008) is available, including
lower effective temperatures, but we have decided to employ Kurucz's models since
they are more consistent with the opacities and equation of state we adopt in
our spectral synthesis calculations.

ATLAS9 models are plane-parallel and in LTE. 
Non-LTE models, for example, the 
BSTAR2006\footnote{Available on 
http:\/\/nova.astro.umd.edu\/Tlusty2002\/BS06-Vispec.html ? for optical 
spectra; and http:\/\/nova.astro.umd.edu\/Tlusty2002\/BS06-UVspec.html 
for ultraviolet spectra.} 
grid by Lanz \& Hubeny (2007), would be more appropriate for the 
warmer stars in the grid, but we made a compromise in this regard in order to 
retain the homogeneity of the model collection.
The adopted ATLAS9 models take the solar reference abundances from Asplund et al. (2005),
and include a fix in the H$_2$O line list that introduces significant differences with
earlier models computed with the same code for stars with effective
temperatures $T_{\rm eff}<4000$~K (M\'esz\'aros et al. 2012). The models
consider variations in carbon and $\alpha$-element abundances. In some of the grids
of synthetic spectra presented here we consider consistent abundance variations in the
$\alpha$-elements, in addition to those in the overall metallicity. The mixing-length
parameter is set to 1.25 times the pressure scale length in all models, and
no overshooting is considered.

\subsection{Continuum opacity and equation of state}
\label{cont}

The spectral synthesis calculations require opacities and an equation of
state that connects temperature and density to gas pressure, and provides 
the number of free electrons.
In our radiative transfer code ASS$\epsilon$T (see Section \ref{rt}), 
these calculations are based on the same routines used in 
Synspec (Hubeny \& Lanz 2000, 2017), accounting for the first 99 atoms
in the periodic table and 338 molecules (Tsuji 1964, 1973,  1976), with 
partition functions from Irwin (1981 and updates).

Bound-free opacity from CH, OH,  H$_2^+$, H$^-$, H I, and the first two ionization
stages of He, C, N, O, Na, Mg, Al, Si, and Ca  is included, with data from the 
Opacity project for carbon and heavier elements 
(Opacity Project Team 1995, 1996, and TOPbase). Absorption
associated with the photoionization of neutral and singly-ionized species is calculated
using cross-sections from the Iron Project (Bautista 1997; Nahar 1995). All the
metal photoionization cross-sections have been smoothed according to the expected
uncertainties, as  described by Bautista et al. (1998). Allende Prieto
et al. (2003a) provided these data for the opacity project ions, and
Allende Prieto et al. (2003b) made some fundamental checks on the performance of
the data, and iron in particular, comparing with the solar spectrum.

Free-free opacity from H$^-$, H$_2^+$, and He$^{-}$ is included. Free-free absorption 
due to metals is considered using a hydrogenic cross-section with the Gaunt factor
set to one, while exact factors are used for H and He II.

\subsection{Line opacity}
\label{line}

Line absorption is considered in detail, accounting for transitions of metals 
included in the lists provided by Kurucz on his website\footnote{http://kurucz.harvard.edu.}, 
in many cases derived from semi-empirical
calculations, but upgraded using higher-quality data from the literature. 
In addition
to the original upgrades made by Kurucz, we replaced the van der Waals damping constants 
for metal lines by those computed by Barklem, Piskunov \& O'Mara (2000a), and  
Barklem \& Aspelund-Johansson (2005), when available.
Molecular lines for the most relevant diatomic molecules are from the extensive compilations
provided by Kurucz, including H$_2$, CH, C$_2$, CN, CO, NH, OH, MgH, SiH, 
and SiO. Our compilation reflects the status of Kurucz's public database around 2007.
We are aware that for some of these molecules there are updates worthy of inclusion, 
and some have already been included in Kurucz's website at the time
of writing. TiO transitions from Schwenke (1998) are included only for the 
coolest models ($T_{\rm eff}<5750$ K).

Level dissolution near the H series limits is included using the treatment
described by Hubeny et al. (1994). Rayleigh (H; Lee \& Kim 2004) and electron (Thomson)
scattering are considered properly, as true scattering, in the radiative transfer.
The damping of H lines by collisions with charged particles (Stark broadening) follows
the treatment by Stehl\'e (1994) and Stehl\'e \& Hutcheon (1999) for the Balmer series, 
or Vidal et al. (1970, 1973) for the Lyman, Paschen \& Brackett series, while self-broadening
for H lines is computed with the tables and codes by Barklem, Piskunov \& O'Mara (2000b)
for the lower transitions of the Balmer series, or Ali \& Griem (1955) for other H lines.

\begin{figure*}
\centering
{\includegraphics[angle=90,width=16cm]{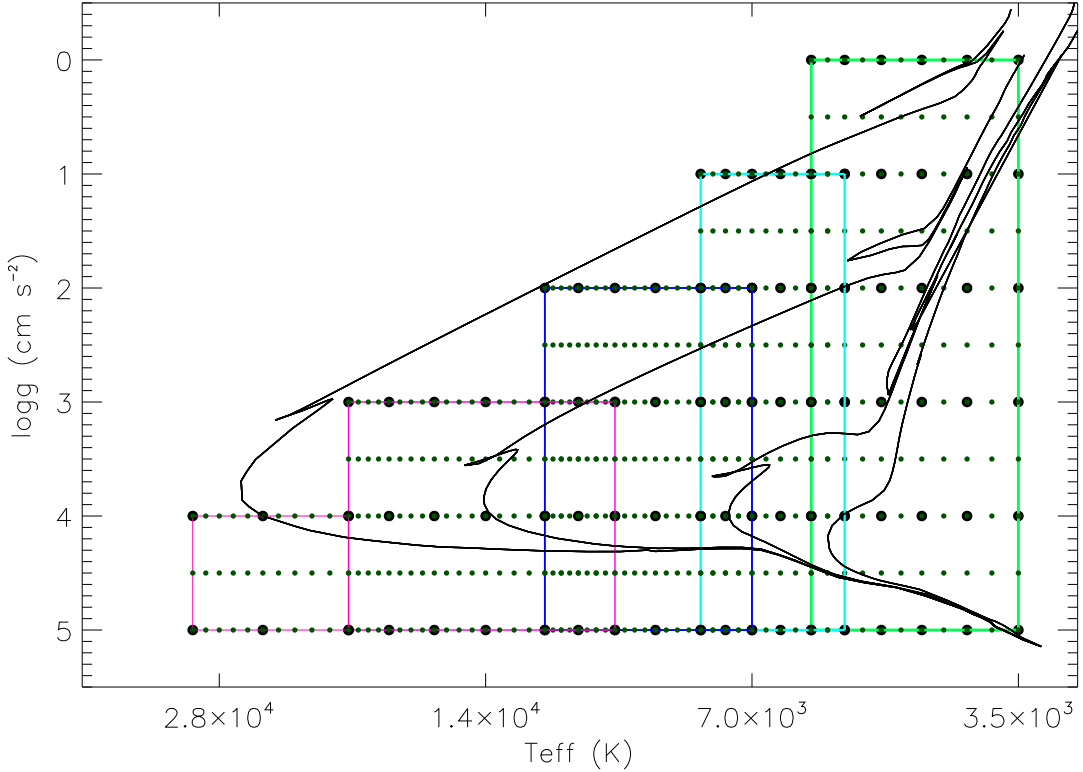}}
\caption{Ranges in $T_{\rm eff}$ and $\log g$ spanned by the libraries. The
metallicity range is always $-5.0$ to $+0.5$. The large dots correspond to the coarse
libraries, while the small ones belong to the finer collection. The colored solid lines
mark the boundaries for each library. Padova isochrones (Girardi et al. 2002, 2004) for
solar compositon and $10^7$, $10^8$, $10^9$ , and $10^{10}$ yr are overlaid.}
\label{lib}
\end{figure*}

\subsection{Radiative transfer}
\label{rt}

All the radiative transfer calculations were performed with 
ASS$\epsilon$T (Koesterke 2009; 
Koesterke et al. 2008) assuming LTE. The code is not public, 
but is available upon request to
one of the authors (LK). Originally developed to perform three-dimensional (3D) radiative transfer, the code
has a powerful one-dimensional (1D) branch that can interpolate on a grid of pre-computed opacities
as a function of density ($\rho$) and temperature ($T$), to substantially speed up the 
calculations. While the calculations included in the coarse grid (see Section \ref{grids})
were done in 'ONE-MOD' mode, 
computing the opacities exactly for each model at every atmospheric depth,  
the  calculations for the finer grids were performed taking
advantage of cubic interpolation of the opacity as a function of $\log \rho$ and $\log T$,
sampling $\rho$ with four points per decade (0.25 dex), and $T$ with 55 points per decade 
(0.018 dex or 250 K at 6000 K).

The  temperatures, mass column densities, and electron densities of the Kurucz models
were respected, avoiding the optional iterative procedure that recomputes the
electron density inside ASS$\epsilon$T. 
For the coarse model grids (see Section \ref{grids}), 
a constant depth-independent 
micro-turbulence velocity of 1.5 km  s$^{-1}$ was adopted. In the finer grids, 
micro-turbulence was varied in constant steps of about 0.3 dex.

For every model the Feautrier algorithm (Feautrier 1963) was adopted in the first
solution of the radiative transfer equation, which provides, using short-characteristics,
the mean intensity at every depth to be used in the calculation of the scattering term. 
The emergent fluxes are later computed from the intensities, obtained using long characteristics, 
for  three inclined angles plus the normal direction chosen for Gauss-Radau quadrature.

The spectral range spans between 118 and 7000 nm, although the grids retained a slightly
smaller one after smoothing and truncation. The frequency steps were chosen independently 
for each model to ensure that they  were never larger than one third of the smallest of the
microturbulence velocity and the thermal Doppler width at the outermost
(and coolest) layer. 
%The continuum is also calculated, with coaser steps constant in
%velocity so that the sampling is 0.02 nm at 300 nm.

\begin{table*}
\label{libpars}
\caption{Parameters covered in libraries.  The number of models included in 
each library is indicated in the column labeled {\it n}. The 'ns' and 'nsc' identifiers are used for
the large and coarse families of libraries, respectively, both at $R=10,000$. 
Similar coarse libraries exist at $R=100,000$ and $R=300,000$, identified as 'hnsc' and
'unsc', respectively.}
\centering
\begin{tabular}{lllrrrr}
\hline
library    &   $T_{\rm eff}$ & $\log g$ & [Fe/H] & [$\alpha$/Fe] & $\log \xi$ & $n$ \\
           &    (K)     &     (cm s$^{-2}$) & &  &     (cm s$^{-1}$)   &   \\
\hline
nsc1  & 3500:6000 (500)  &  0.0:5.0 (1.0) &   $-5$:$+0.5$  (0.5) & 0.5 at [Fe/H]$\le-1.5$, & 0.176 & 432 \\
      &                  &                &               & 0.0 at [Fe/H]$\ge 0$,   &   \\
      &                  &                &               & linear in between   &   \\      
nsc2  & 5750:8000 (500) &  1.0:5.0 (1.0) &      $-5$:$+0.5$  (0.5)  & ---''---  & 0.176   & 360  \\
nsc3  & 7000:12000  (1000) &  2.0:5.0 (1.0) &   $-5$:$+0.5$  (0.5)  & ---''---  & 0.176 & 288  \\
nsc4  & 10000:20000  (2000) &  3.0:5.0 (1.0) &  $-5$:$+0.5$ (0.5)   & ---''---  & 0.176 & 216   \\
nsc5  & 20000:30000 (5000) &  4.0:5.0 (1.0) &   $-5$:$+0.5$  (0.5)  & ---''---  & 0.176  & 72  \\
\hline
ns1  & 3500:6000   (250) &  0.0:5.0 (0.5) &   $-5$:$+0.5$ (0.25)  & $-1$:$+1$ (0.25) &  $-0.301$:$+0.903$ (0.301) & 136,125 \\
ns2  & 5750:8000  (250) &  1.0:5.0 (0.5) &   $-5$:$+0.5$ (0.25) & $-1$:$+1$ (0.25) &  $-0.301$:$+0.903$ (0.301) & 101,250  \\
ns3  & 7000:12000 (250) &  2.0:5.0 (0.5)  &   $-5$:$+0.5$ (0.25) & $-1$:$+1$  (0.25) &  $-0.301$:$+0.903$ (0.301) & 86,625  \\
ns4  & 10000:20000 (500)   &  3.0:5.0 (0.5) &   $-5$:$+0.5$ (0.5) & $-1$:$+1$ (0.5) &  $-0.301$:$+0.903$ (0.301) & 61,875 \\
ns5  & 20000:30000 (1000)  &  4.0:5.0 (0.5) &   $-5$:$+0.5$ (0.5) & $-1$:$+1$ (0.5) &  $-0.301$:$+0.903$ (0.301) & 37,125 \\
\hline
\end{tabular}
\end{table*}

\begin{figure*}
\centering
{\includegraphics[angle=90,width=16cm]{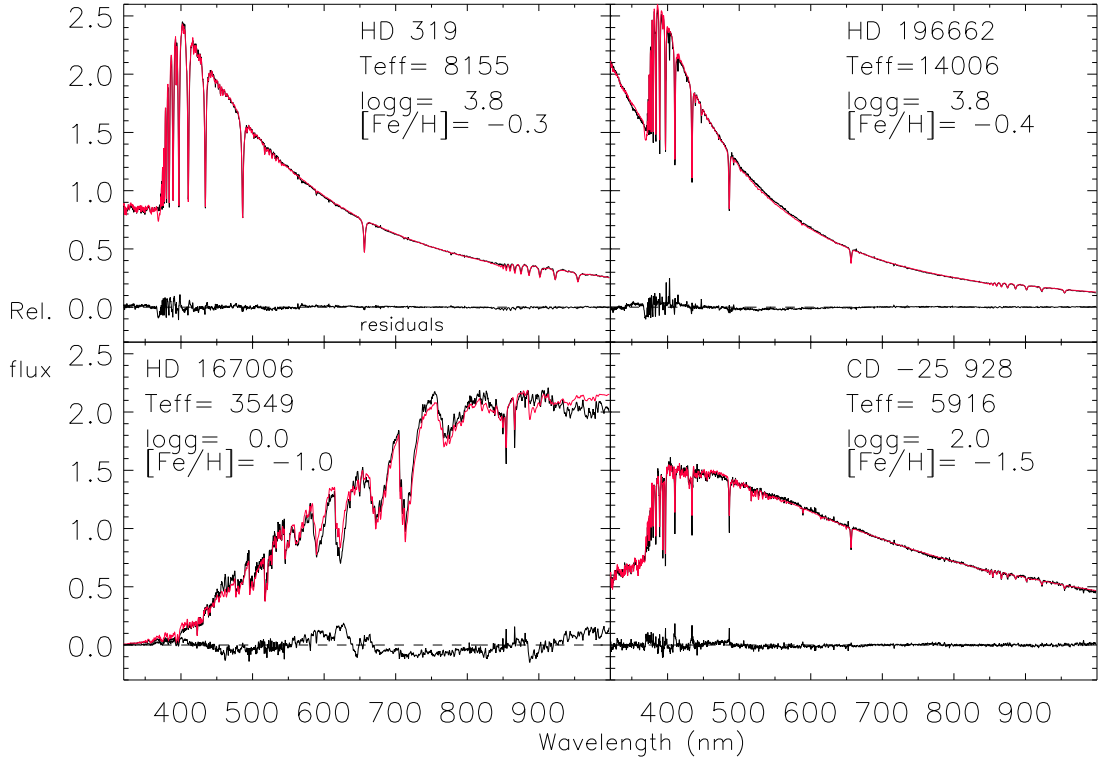}}
\caption{Comparison between STIS observations from NGSL (black lines) 
and best-fitting models (red). The residuals are also shown in black. The labels show the parameters 
recovered with FERRE. The vertical scale corresponds to the stellar flux 
(erg cm$^{-2}$ s$^{-1}$ \AA$^{-1}$) normalized by 
the mean value in the selected spectral range.}
\label{mdl}
\end{figure*}

\section{Spectral grids}
\label{grids}

As the effective temperature of the models increases, radiation pressure makes them 
unstable at low gravity. Accordingly, the smallest value in the $\log g$ range is
progressively reduced in the construction of model atmospheres (see Fig. 2 
in M\'esz\'aros et al. 2012), a feature that our library inherits. 

Following the same strategy as in the APOGEE spectral libraries (Zamora et al. 2013; 
see also Allende Prieto 2009),
we split our spectral model collection into five libraries-files according to
effective temperature, one for the range
3500--6000 K ($\log g \ge 0$), a second for the range 5750--8000 K
 ($\log g \ge 1$), a third for 7000--12,000 K ($\log g  \ge 2$),
a fourth for 10,000--20,000 K ($\log g \ge 3$), and a fifth for 
20,000--30,000 K ($\log g \ge 4$).

A set of coarse libraries with a resolution of $R\equiv \lambda/\delta\lambda = 10,000$ is
available, and identical files corresponding to $R=100,000$ and  $R=300,000$
are available as well.\footnote{
The libraries are available from the CDS and at ftp://carlos:allende@ftp.ll.iac.es/collection.}
These only consider three atmospheric parameters ($T_{\rm eff}$, $\log g$, and [Fe/H]), and
have steps in the parameters that increase with effective temperature: 500 K for the first
and the second libraries ($3,500 \le T_{\rm eff} \le 8000$ K), 1000 K for the third, 
2000 K for the fourth, and 5000 K for the fifth and warmer. All span the metallicity
range $-5 \le$[Fe/H]$\le +0.5$. These libraries cover the spectral range between 
120 and 6500 nm, sampling the spectra with equidistant steps in $\log \lambda$; for 
the $R=10,000$ grids the step size is $1.434\times 10^{-5}$, 
equivalent to $\sim 10$ km s$^{-1}$.

These coarse libraries are useful for some applications, but others will need 
finer resolution in the parameters. A second set of libraries retains the 
subdivisions by $T_{\rm eff}$, and they span the same range in $T_{\rm eff}$, 
$\log g$\ and [Fe/H], but have much finer steps in the parameters, and 
include additional dimensions, namely
microturbulence (parameterized on a logarithmic scale $\log_{10} \xi$) and the 
[$\alpha$/Fe] abundance ratio. Needless to say, the size of these later libraries-files  
is much larger than the coarse ones, between a few and tens of gigabytes. Figure \ref{lib}
illustrates the ranges in the plane $T_{\rm eff}$-$\log g$ of the libraries.
The parameter ranges and step sizes are given in Table 1. These finer 
libraries are provided only at $R=10,000$, with a spectral range between 200 and 2500 nm,
 and the same  equidistant steps in $\log \lambda$ as the smaller libraries.
%and at $R=100,000$, with a spectral range between 350 and 1000 nm.?
All the libraries give the Eddington flux (first-moment of the radiation field) 
$H_\lambda$ ($\equiv  F_{\lambda}/(4 \pi)$) at 
the stellar surface in units of erg cm$^{-2}$ s$^{-1}$ \AA$^{-1}$.

\begin{table}
\label{ngsl}
\caption{Parameters for several stars with observations from the Next Generation Spectral
Library}
\begin{tabular}{llllrrr}
\hline
        & \multicolumn{3}{c}{This work} & \multicolumn{3}{c}{NGSL} \\
star    &   $T_{\rm eff}$ & $\log g$ & [Fe/H] & $T_{\rm eff}$ & $\log g$ & [Fe/H] \\
%        &    (K)     &     (cm s$^{-2}$) & & (K)     &     (cm s$^{-2}$) & \\
\hline
HD 167006  & 3549  &    0.0 &   -1.0 &   3536  &  0.4 & -0.3 \\
CD -25 928 &  5916 &    2.0 &   -1.5 &   6426  &  3.0 & -1.2 \\
HD 319     & 8155  &    3.8 &   -0.3 &   8195  &  3.9 & -0.4 \\
HD 196662 & 14006 &   3.8 &   -0.4 &   14204 &  3.7 & -0.6 \\
\hline
\end{tabular}
\end{table}

\begin{figure*}
\centering
{\includegraphics[angle=0,width=8cm]{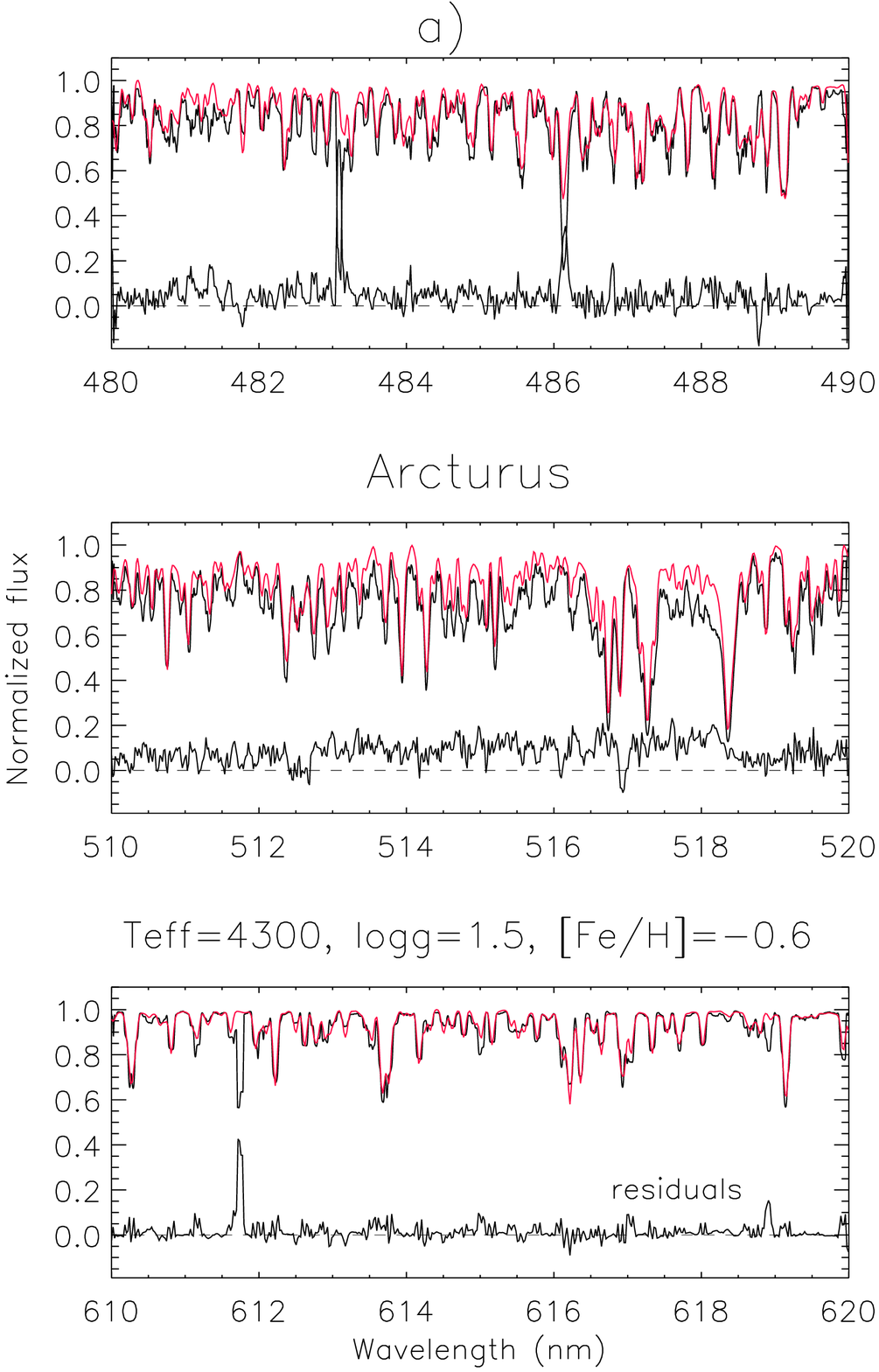}}
{\includegraphics[angle=0,width=8cm]{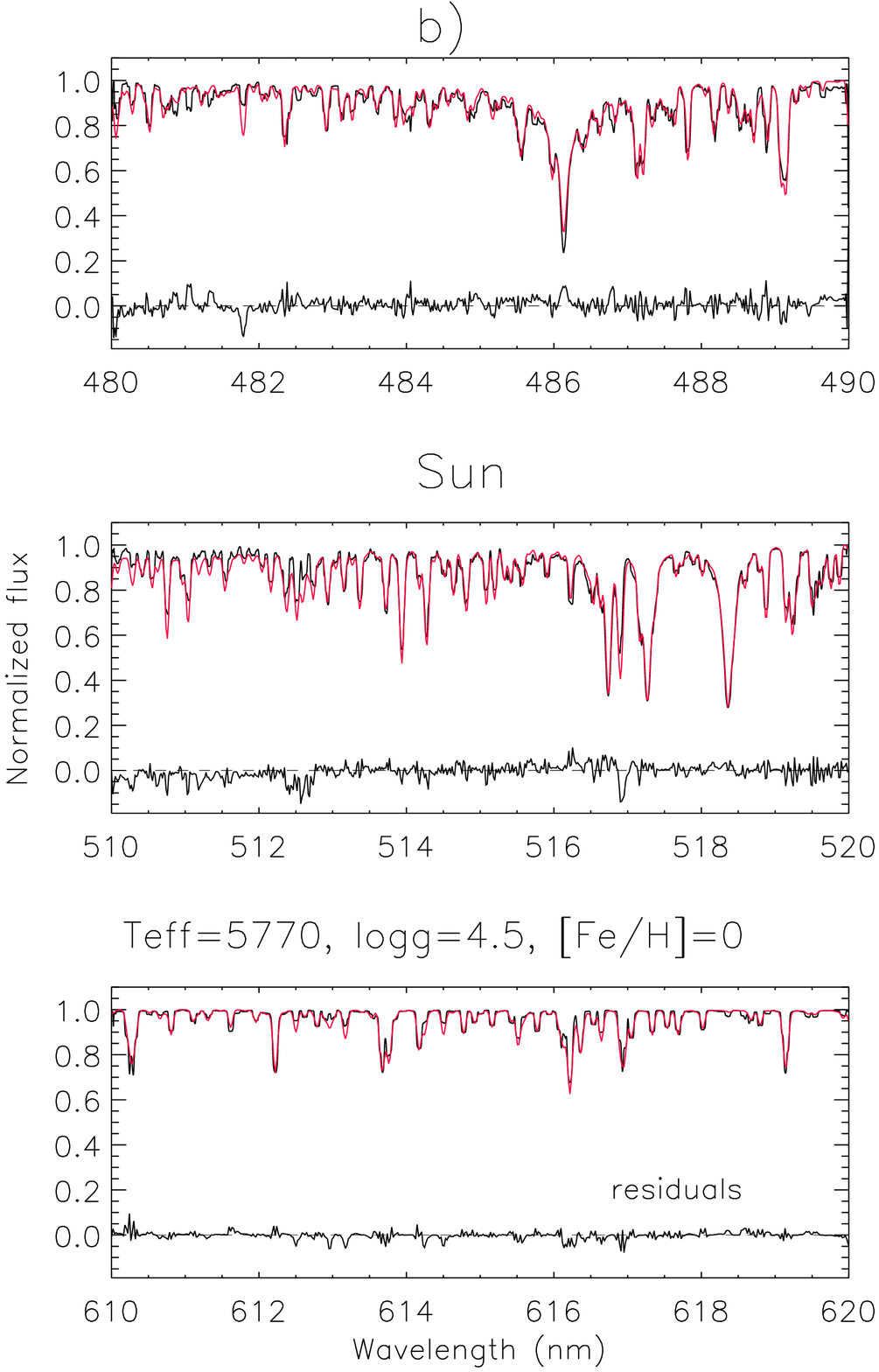}}
{\includegraphics[angle=0,width=8cm]{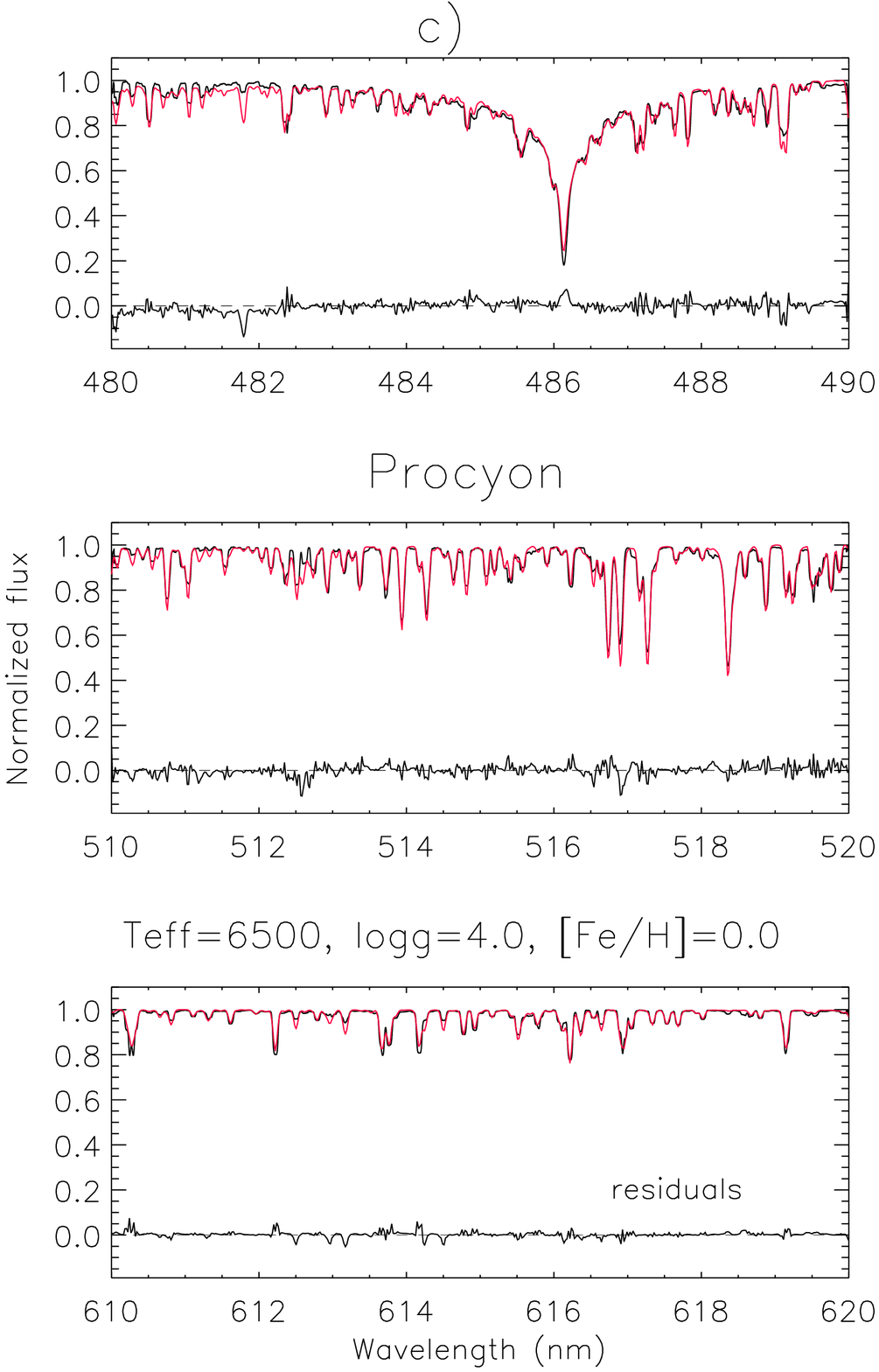}}
{\includegraphics[angle=0,width=8cm]{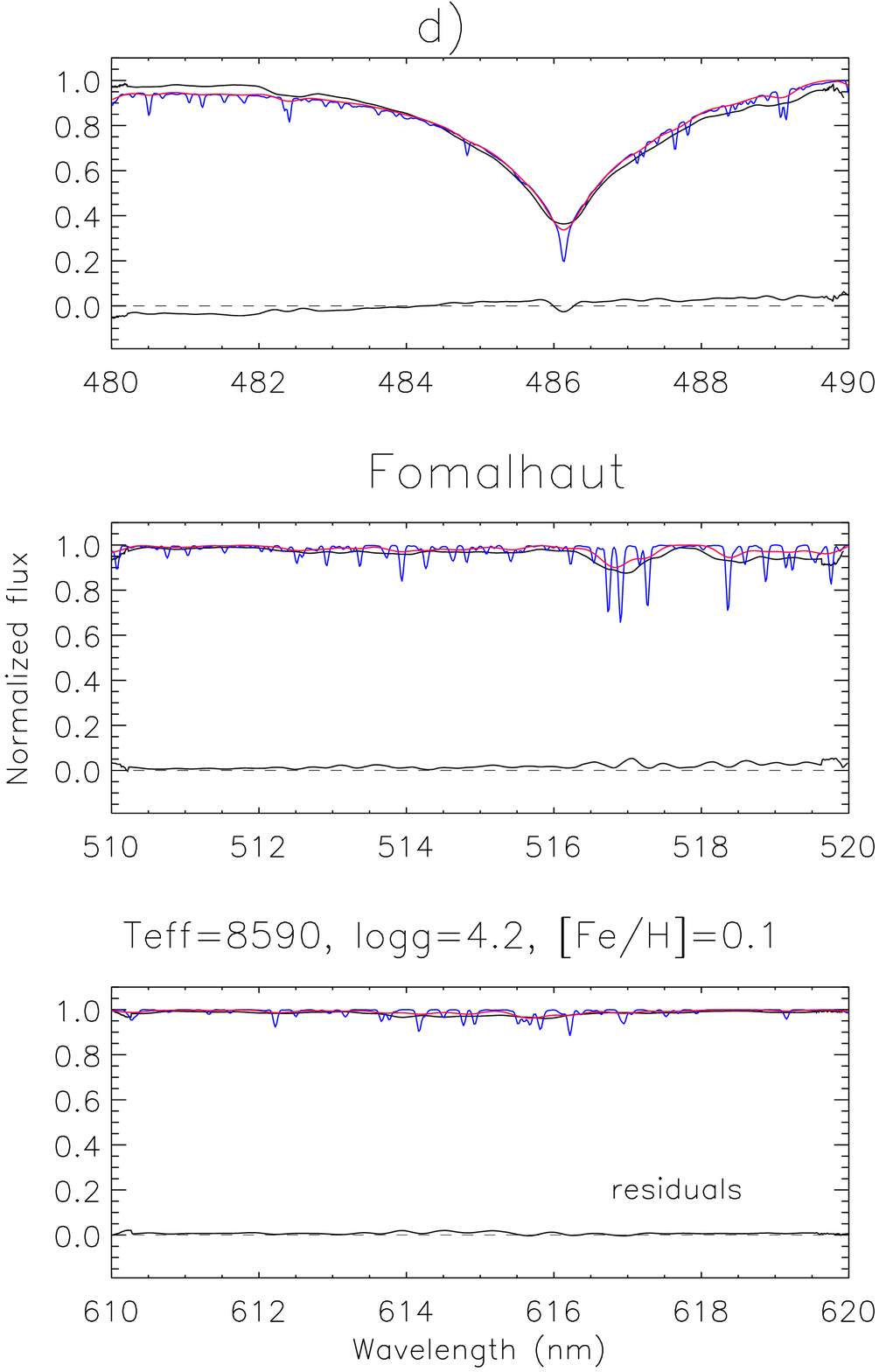}}
\caption{Comparison between high-resolution observations of a) Arcturus, b) the Sun, c) Procyon, 
and d) Fomalhaut (black lines) 
with interpolated models for their nominal atmospheric parameters (red). The residuals are 
also shown in black. 
The labels show the parameters adopted in the comparison. In the case of Fomalhaut, 
the effect of rotational broadening is obvious, and the interpolated
model is shown before (blue) and after smoothing (red).}
\label{high}
\end{figure*}

\section{Comparison with observations}
\label{data}

A library of model spectra without an assessment of how well it matches reality is useless.
We therefore provide in this section several examples of how well the library performs
in comparison with a selection of observations. We have chosen to compare the models
with data from the second version of the Next Generation Spectral Library (Gregg et al. 2006; 
Lindler \& Heap 2008;  Heap \& Lindler 2016), which contains low-resolution
flux-calibrated observations from the  STIS instrument onboard the Hubble Space
Telescope. 

We selected four representative stars from the library, spanning a reasonable range 
in $T_{\rm eff}$, and fitted their STIS spectra with our libraries (the five-parameter version)
and the FERRE code (Allende Prieto et al. 2006), which finds by optimization (interpolating
in the libraries) the set of parameters that leads to the minimum of the $\chi^2$ statistics. 
The library spectra were smoothed to a resolving
power of 510 (Bohlin 2007, Allende Prieto \& del Burgo 2016). 
The results are shown in Table 2 and
illustrated in Fig. \ref{mdl}. With the exception of CD $-25$ 928, the agreement between the 
two sets of effective temperatures is excellent, as one would expect since both analyses 
fit the same spectral energy distributions. The disagreement for CD $-25$ 928 is likely related
to a difference in surface gravity, which in our analysis is obtained as well from the
STIS spectrum, while those provided with the NGSL are tied to the trigonometric parallaxes
from Hipparcos. The metallicities show fair agreement.

\begin{figure}
\centering
{\includegraphics[angle=0,width=9cm]{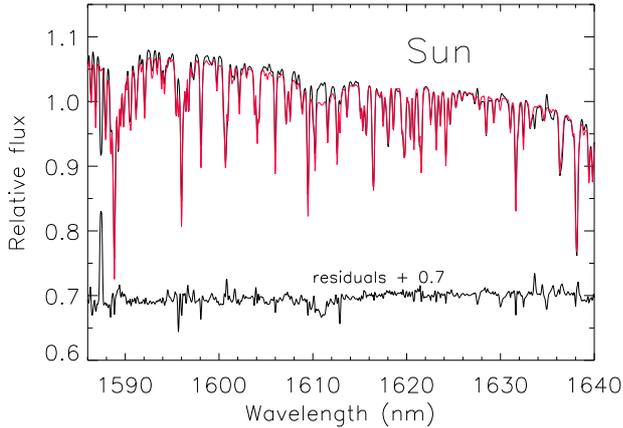}}
\caption{Comparison between the APOGEE spectrum of asteroid Vesta (black)
and a model interpolated for the solar parameters (red). The residuals are 
also shown in black.}
\label{vesta}
\end{figure}

Figure \ref{high} illustrates the comparison between the library and the high resolution spectra
of four stars with well-known parameters at high resolving power ($R=20,000$). We have chosen
three important regions widely used in the literature: the vicinity of H$\beta$, the region 
around the Mg I$b$ triplet, and the band near the Ca I $\lambda$616.2 nm transition. The
parameters for Arcturus were adopted from Ram\'{\i}rez \& Allende Prieto (2011; data from 
Hinkle et al. 2000), those for
Procyon from Allende Prieto et al. (2002; data from Griffin \& Griffin 1979), and those 
for Fomalhaut from Mamajek (2002) and 
G\'asp\'ar et al. (2016; data from Allende Prieto et al. 2004). The solar spectrum is
from Kurucz (2005). The main features present in the spectra are captured by the models.

A section of the APOGEE spectrum of solar light reflected off Vesta, 
smoothed to $R\simeq 10,000$, is compared to a spectrum interpolated for the solar 
parameters in Fig. \ref{vesta}. The agreement between data and models in 
this region is nearly as good as that 
found in the optical, although the presence of a few missing transitions in the 
models and residuals caused by imperfectly subtracted OH sky lines are obvious. 
We note that the H line at 1611 nm is significantly stronger in the model, that is, 
the situation is the opposite to the case of H$\beta$, shown in Fig. \ref{high}b.
Both the observed and model spectra retain their original slope; they have not
been continuum normalized but simply divided by their mean values in the selected region.
Further comparisons of these models with observations can be found, for example, in 
Aguado et al. (2017, 2018), Allende Prieto (2016), Allende Prieto \& del Burgo (2016),  
La Barbera et al. (2017), R{\"o}ck et al. (2017), or Yoakim et al. (2017).

\section{Summary and conclusions}
\label{sum}

The data that enter the construction of model atmospheres and the calculation 
of synthetic stellar spectra are in constant upgrade, and so are the modeling tools. 
In this paper we describe a set of calculations that represent our own efforts to 
improve existing models over more than a decade, and which therefore are, in some ways, 
already outdated. Nonetheless, these models provide significant advantages over a number
of other existing libraries, most notably an extended wavelength coverage and the
inclusion of warmer models, and constitute a collection that can be used 
for practical analysis of stellar spectra or building stellar population models.

The model collection is provided as a set of five separate libraries, each spanning
a regular grid in the parameter space. We publish models in which only three parameters are
considered ($T_{\rm eff}$, $\log g$, and [Fe/H]), with a constant micro-turbulence of 
{\bf 1.5} km s$^{-1}$, and an abundance ratio [$\alpha$/Fe] that depends on [Fe/H]. We also 
present additional model libraries in which 
five parameters are changed ([$\alpha$/Fe] and micro-turbulence, in addition to the
previous three). All the libraries are in a format ready to use with the 
FERRE code\footnote{Available from github.com/callendeprieto/ferre}. This format is 
straightforward and has been described in the FERRE documentation.

A comparison with several spectra from the NGSL 
shows that the models capture the main spectral features and the parameters inferred from 
the observations are in good agreement with those previously derived by the makers of the NGSL.
An in-depth analysis of this library with these models, as well as of the MILES library 
(S\'anchez-Blazquez et al. 2006; Cenarro et al. 2007), will be the subject of future work.
The library provides also a reasonable description of stellar spectra at high
resolution, as checked against spectra of stars as cool as the K giant Arcturus, 
or as warm as the A-type star Fomalhaut, and will be systematically used in the analysis
of observations from instruments such as SONG (Grundahl et al. 2017) or ESPRESSO (Pepe et al. 2014). 

\begin{acknowledgements}
CAP's research is partially funded by the Spanish MINECO under grant 
AYA2014-56359-P. PSB acknowledges financial support from the Swedish 
Research Council  and the project grant ``The New Milky Way'' from the 
Knut and Alice Wallenberg Foundation. MB acknowledges support from the 
US National Science Foundation Astronomy and Astrophysics Program (AST- 1313265). 
SNN acknowledges the support of DOE grant DE-FG52-09NA29580 and NSF grant AST-1312441.
We are indebted to the referee of the paper for multiple useful suggestions
that improved the quality of the presentation, and to the A\&A language editor, Ruth Chester, 
for an excellent job.

\end{acknowledgements}

\end{document}